%% file: Cracow_proc.tex
\documentstyle[12pt,epsfig,bezier]{article}

\renewcommand{\cal}{\mathcal}
\def\Journal#1#2#3#4{{#1}{\bf #2} (#4) #3}

\def\CPC{\em Comput. Phys. Commun.~}
\def\FdP{\em Fortschritte der Physik~}

\def\PLB{{\em Phys. Lett.} {\bf B}}
\def\PRD{{\em Phys. Rev.} {\bf D}}
\def\ZPC{{\em Z. Phys.} {\bf C}}
\def\JPG{{\em J. Phys.} {\bf G}}
\include{lk_newcommand}

\newcommand {\ql} {\mbox{$Q^2_{l}  $}}

\newcommand {\qh} {\mbox{$Q^2_{h}  $}}

\newcommand {\oa}  {${\cal O}({\alpha})$}

\newcommand {\oaal} {${\cal O}(({\alpha}L)^2)$}

\begin{document}
\begin{titlepage}
\begin{flushleft}
{DESY 96-243}\\
{\tt hep-ph/9611426}\\
{November 1996}
\end{flushleft}
\vspace*{1.5cm}
\normalsize

\begin{center}
{\Large {\bf QED and Electroweak Corrections to Deep Inelastic Scattering}}

\vspace{0.25cm}

{\large 
Dima Bardin$^{a,b}$,~Johannes Bl\"umlein$^a$,~Penka Christova$^c$,}

\vspace{0.15cm}

{\large
Lida Kalinovskaya$^b$ and Tord Riemann$^a$ }

\vspace{1.0cm}
\noindent
{\it $^a$ DESY--Zeuthen,} \\
{\it Platanenallee 6, D--15735 Zeuthen, Germany }\\

\vspace{0.1cm}
\noindent
{\it $^b$ Bogoliubov  Laboratory for Theoretical Physics, JINR,} \\
{\it ul. Joliot-Curie 6, RU--141980 Dubna, Russia } \\

\vspace{0.1cm}
\noindent
{\it $^c$ Bishop Konstantin Preslavsky University of Shoumen,} \\
{\it 9700 Shoumen, Bulgaria   } \\

\end{center}

\vspace{1.0cm}

\normalsize
\centerline{\bf Abstract}

\noindent
We describe the state of the art in the field of
radiative corrections for deep
inelastic scattering.
Different methods of calculation of radiative corrections are reviewed.
Some new results for QED radiative corrections for polarized
deep inelastic scattering at HERA
are presented.
A comparison of results obtained by the
codes {\tt POLRAD} and {\tt HECTOR} is given
for the kinematic regime
of the HERMES experiment.
Recent results on radiative corrections to deep inelastic scattering with 
tagged photons are briefly discussed.

\vspace{.5cm}
 
\centerline{\small Contribution to the Proceedings of the 3rd
  International Symposium}

\centerline{on Radiative Corrections, Cracow, Poland, August 1-5, 1996}

\end{titlepage}
				

\newpage

\begin{center}
{\Large {\bf QED and Electroweak Corrections to Deep Inelastic Scattering}}

\vspace{1cm}

\begin{large}

{\underline {Dima Bardin}}$^{a,b}$\footnote{Speaker at CRAD'96 Symposium.},
Johannes Bl\"umlein$^{a}$, 
Penka Christova$^{c}$\footnote{Supported by PECO contract ERBCIPDCT-94-0016.},
\\
Lida Kalinovskaya$^{b}$ and Tord Riemann$^{a}$\\
 
\end{large}

\vspace{.5cm}

$^a$ DESY--Zeuthen,
     Platanenallee~6,~D-15735~Zeuthen, Germany\\
$^b$ Bogoliubov Laboratory for Theoretical Physics, \\
JINR, ul. Joliot-Curie 6, RU--141980 Dubna, Russia \\
$^c$ Bishop Konstantin Preslavsky University of Shoumen, \\
9700 Shoumen, Bulgaria    \\
\end{center}
\begin{quotation}
\noindent
{\bf Abstract:}
\noindent
We describe the state of the art in the field of
radiative corrections for deep
inelastic scattering.
Different methods of calculation of radiative corrections are reviewed.
Some new results for QED radiative corrections for polarized
deep inelastic scattering at HERA
are presented.
A comparison of results obtained by the
codes {\tt POLRAD} and {\tt HECTOR} is given
for the kinematic regime
of the HERMES experiment.
Recent results on radiative corrections to deep inelastic scattering with 
tagged photons are briefly discussed.

\end{quotation}

\section{Introduction}
\label{sect1}

The knowledge of QED, QCD, and electroweak (EW) radiative corrections (RC)
to the different deep inelastic scattering (DIS) processes 
is indispensable
for the precise determination of the nucleon structure functions (SF).	
The forthcoming high statistics measurements of unpolarized and polarized SF
at H1, ZEUS, HERMES, and SLAC require the knowledge of the RC
at the percent level. This has to be met by adequately precise theoretical 
calculations.

In this report we summarize the actual status
in the field of RC's for DIS. 
In section 2, we present a short review of different methods used
with an emphasis on the so-called Model Independent approach (MI).
Here, we also describe results of a
recent new calculation~\cite{ph} of the QED corrections for polarized DIS 
including both $\gamma$ and $Z$-boson exchange 
and accounting for all twist-2 contributions
to the polarized SF's 
for both
longitudinally and transversely polarized nucleons.
In section 3, we present some new numerical results of this 
calculation. 
Section 4 sketches briefly recent results on the RC's for DIS with
tagged photons~\cite{tp}.
This report represents a natural continuation of a talk~\cite{db} presented
at the Warsaw Rochester Conference.
In that talk an additional motivation is presented
showing why the field is still a very vivid one.

\section{Different Approaches}

\subsection{A Qualitative Comparison of Monte Carlo,
Semi-Analytic, and Deterministic Approaches}

Until recently, two basic approaches to the RC's for DIS were used:
\begin{itemize}
\item The Monte Carlo (MC) approach aims at the construction of
precise event generators (MCEG). This approach is exclusive and
deals with completely differential cross-sections. Therefore
it is rather flexible with respect to experimental applications,
e.g. allowing for cuts. MCEG are real tools for data analysis.
In principle, this approach suffers of statistical errors
although a very impressive performance of MCEG's
has been reached in recent years, see~\cite{Bhabha} and~\cite{LEP2}.
Typical examples of MCEG's for DIS are: {\tt HERACLES}~\cite{hr},
{\tt LESKO-F}~\cite{lf}, and {\tt KRONOS}~\cite{kr}.

\item Semi-Analytic (SAN) approaches aim at partly
integrated cross-sections. Therefore they are much less flexible 
concerning possible cuts
as compared to MCEG's.
Only a limited number of inclusive distributions can be usually 
evaluated and no event generation is possible.
However, the method provides
fast and precise codes, provide exact
benchmarks for MCEG's.
The underlying physics is clearly exhibited, and
sometimes appealing formulae emerge as a reward. 
These are the reasons why people will
probably always try to perform SAN calculations. 
Furthermore, SAN codes may be used for fitting 
of theory predictions to experimental data
at the final phase of their analysis.
Examples of SAN codes for DIS are: {\tt HELIOS}~\cite{hl},
{\tt TERAD91}~\cite{tr}, 
{\tt FERRAD}~\cite{fr,mots},
{\tt APHRODITES}~\cite{ar}, {\tt POLRAD}~\cite{pr} and finally 
{\tt HECTOR}~\cite{he}, to which this talk is largely related.
\end{itemize}

Recently people began to use the so-called
{\it Deterministic Approach} (DA), see, for instance,~\cite{da}.
The DA is an alternative to the MC approach but without the ability of
event generation.
It also operates with
completely differential cross-sections but integrates avoiding MC methods.
A faster computing than in the case of MC may emerge since the integration
is based on methods possessing better convergency
(see the talk by T.~Ohl~\cite{to} in these proceedings for a discussion of
basic issues of DA).
A necessary feature of DA should be the access to
{\bf any} realistic experimental cuts.
This is usually achieved 
by the explicit solution of the relevant kinematic inequalities
for the phase space boundaries.
The elements of the DA are used in two of our recent codes,
${\mu}${\bf e}{\it la}~{\tt 1.00}~\cite{mu} and
one of the new branches in {\tt HECTOR 1.11}~\cite{h2}.
\subsection{Model Independent Approach}
The Model Independent approach to the problem under consideration
is usually understood as the description of the
QED RC's to {\bf only} the leptonic line of the
Born-level Feynman diagrams. The hadronic part 
of the diagrams is assumed to be untouched.
Therefore both the Born approximation and the radiative diagrams
contain the same {\em hadronic tensor} accessing
hadron dynamics through a potentially {\em Model Independent} description 
by means of the {\em structure functions}. This is possible only
for the neutral current (NC) DIS 
where a continuous flow of the electric charge through the
leptonic line ensures the QED gauge invariance of the description
to all orders. The MI approach was comprehensively reviewed
in~\cite{mi} recently.

\subsubsection{Born cross-section for the process $ep \rightarrow e X$}
Here we present a complete set of formulae for the polarized DIS Born
cross-section which can be written as a
contraction of leptonic ($L^{\mu\nu}$) and
hadronic ($W_{\mu\nu}$) tensors:
\ba
d\sigma_{_{\mr BORN}}
 =  \frac {2\pi \alpha^2}{Q^4}
           y
 \Biggl[ L^{\mu\nu} W_{\mu\nu} \Biggr]
  dx dy,
\label{sborn2}
\ea
with the usual notation for momentum transfer ($q$),
the invariants ($Q^2,S$) 
\ba
q=k_1-k_2, \qquad  Q^2 = -q^2=-t ,  \qquad  S = 2 (p.k_1), 
\label{minv}
\ea
and the Bjorken scaling variables ($x,y$)
\ba
x = \frac{Q^2}{S y}, \qquad   
y = \frac{p.q}{p.k_1}.
\label{xyscal}
\ea
\noindent
The polarization of the lepton beam is described by
the spin density matrix
\ba
\rho(k_1)= \sum_s u^{\small s}(k_1) {\bar u}^{\small s}(k_1) = \frac{1}{2}
  \left( 1 - \gamma_5 \hat \xi_e \right)
           \left( \hat{k_1} + m \right),
\label{spdenm}
\ea
where $\xi_e$ is the lepton polarization vector, satisfying
\ba
 \xi_e.k_1 =0.                 
\ea
The {\bf leptonic tensor} on the Born-level is derived
straightforwardly 
\ba
 L^{\mu\nu} &=&
     2\left[k^{\mu}_1 k^{\nu}_2+k^{\nu}_1
 k^{\mu}_2- g^{\mu \nu}(k_1.k_2)  \right]
   L_{_S} (Q^2)
\nll &&
 +2 i \Pe k_{ 1 \alpha} k_{ 2 \beta} \varepsilon^{\alpha\beta\nu\mu}
   L_{_A} (Q^2).
\label{lnrt1}
\ea
It contains a symmetric (S) and an antisymmetric (A) part
\ba
   L_{_S} (Q^2) &=& Q^2_{e} + 2|\Qe|
               \left( v_e - \Pe \lambda_e a_e  \right) \chi(Q^2)
\nll
&&    + \left( v^2_e + a^2_e - 2 \Pe  \lambda_e v_e a_e \right) \chi^2(Q^2),
\nll
   L_{_A} (Q^2)&=&
       -\Pe \lambda_e
        Q^2_e + 2 |Q_e|
        \left( a_e -\Pe \lambda_e v_e \right) \chi(Q^2)
\nll
&& +\Biggl(2 v_e a_e - \Pe\lambda_e \left( v^2_e + a^2_e\right)
             \Biggr) \chi^2(Q^2).
\label{lnrt2}
\ea
In~(\ref{lnrt2}), $v_e$ and $a_e$ stand for the vector and axial-vector
couplings of electrons to $Z$-boson, 
 $\Pe = 1$  for a particle beam  and
 $\Pe =-1$  for an antiparticle beam, $\chi (Q^2)$ is
the $\gamma/Z$ propagator ratio
\ba
 \chi (Q^2) =
{G_\mu \over\sqrt{2}}{M_{Z}^{2} \over{8\pi\alpha}}{Q^2 \over
{Q^2+M_{Z}^{2}}}.
\label{chi}
\ea

The expression~(\ref{lnrt1}) possesses a nice factorization property
when the tensorial structures decouple from $\gamma$ and $Z$ propagators
and couplings.
This is a consequence of the  
{\it Ultra Relativistic Approximation} (URA) for
the longitudinal polarization of incoming leptons which
implies
\ba
\xi_e =\frac{\lambda_e}{m}k_1.       
\label{xiura}
\ea

This approximation is very accurate for the description of the Born
cross-section since it results in neglection of terms of
${\cal O}(m^2/Q^2)$.
It is not precise enough, however,
for the description of radiative polarized DIS, and as a result
the factorization property~(\ref{lnrt1})
is lost.

The {\bf hadronic tensor} is being constructed from general 
principles of invariance (Lorenz invariance, current conservation). 
There is no
unique presentation for it in the literature. We use the form of
ref.~\cite{blukoch}, where one can also find a review of other
presentations used:
\ba
\label{hadten}
 W_{\mu\nu}     &=& p^0 (2\pi)^6 \sum \int
          \langle p^{'}| {\cal J}^{^{J_1}}_{\mu} |p     \rangle
          \langle p    | {\cal J}^{^{J_2}}_{\nu} |p^{'} \rangle 
          \delta^4 (\sum_{i} p^{'}_i- p^{'}) \prod_{i} d p^{'}_i
\nll
  &=& \left(-g_{\mu\nu}+ \frac{q_\mu q_\nu }{q^2} \right)
                                             F^{^{J_1 J_2}}_1(x,Q^2)
 +\frac{\widehat{p_\mu}\widehat{p_\nu}}{p.q}
                                             F^{^{J_1 J_2}}_2(x,Q^2)
\nll
  &&   - i e_{\mu\nu\lambda\sigma} \frac{q^\lambda p^\sigma}{2 p.q}  
                                             F^{^{J_1 J_2}}_3(x,Q^2)       
     + i e_{\mu\nu\lambda\sigma}\frac{q^\lambda {\pspin}^\sigma} {p.q}
                                             g^{^{J_1 J_2}}_1(x,Q^2)
\nll
  && + i e_{\mu\nu\lambda\sigma}
          \frac {q^{\lambda}(p.q {\pspin}^\sigma
                      -{\pspin}.q p^\sigma)}{(p.q)^2}
                                             g^{^{J_1 J_2}}_2(x,Q^2)      
\\
  && +      \left[\frac{\widehat{p_\mu} \widehat{{\pspin}_\nu}
        +   \widehat {{\pspin}_\mu} \widehat{p_\nu}}{2}
                -  {\pspin}.q
         \frac{\widehat{p_\mu}\widehat{p_\nu}}{p.q}\right]\frac{1}{p.q} 
                                             g^{^{J_1 J_2}}_3(x,Q^2)
\nll
  && +   {\pspin}.q \frac{\widehat{p_\mu} \widehat{p_\nu}}{(p.q)^2}
                                             g^{^{J_1 J_2}}_4(x,Q^2)
               +  \left( - g_{\mu\nu}+ \frac{q_\mu q_\nu }{q^2}\right)
                  \frac{ {\pspin}.q  }{p.q}       
                                             g^{^{J_1 J_2}}_5(x,Q^2),
\nonumber
\ea
with
\ba
 \widehat {p_\mu}   =   p_\mu - \frac{p.q }{q^2}  q_\mu,\qquad
 \widehat {{\pspin}_\mu}  =  {\pspin}_\mu 
- \frac{{\pspin}.q}{q^2}  q_\mu,
\label{haddef}
\ea
and $\pspin$ is the four-vector of the nucleon spin.
In the nucleon rest frame one has
\ba
    s= \lambda_p M (0,{\vec n} ).
\label{spin}
\ea
The hadronic
structure functions  $F^{^{J_1 J_2}}_i$ and  $g^{^{J_1 J_2}}_i$
are associated with the respective currents $J_1,J_2=\gamma,Z$.

Contracting the leptonic and hadronic tensors in~(\ref{sborn2}), one
derives the three Born cross-sections,
depending on the nucleon spin orientation.
The unpolarized DIS Born cross-section reads
\ba
\frac{\ds d\sigma^{^{U}}_{_{\mr BORN}}}{\ds dx dy}
 =    \frac{2\pi \alpha^2 }{Q^4}
\sum_{i=1}^{3} \;  S^{^U}_{i}(y,Q^2) {\cal F}_i(x,Q^2),
\label{unpol}
\ea
with the kinematical factors
\ba
  S^{^U}_{1}(y,Q^2) &=&  2y Q^2,                           \nll
  S^{^U}_{2}(y,Q^2) &=&   2 \left[S (1-y)-xy \Mp\right],     \nll
  S^{^U}_{3}(y,Q^2) &=&   (2 - y) Q^2.
\label{sunpol}
\ea
The two polarized DIS
 Born cross-sections are
\ba
\frac{\ds d\sigma^{^{L,T}}_{_{\mr BORN}}}{\ds d\xl dy}
&=& \frac{2 \pi \alpha^2}{Q^4} f^{^{L,T}}
     \sum_{i=1}^{5} \;  S^{^{L,T}}_{gi}(y,Q^2) {\cal G}_i(x,Q^2),
\label{polar}
\ea
where
\ba
f^{^{L}}&=&\lambda^{^{L}}_p,
\nll
f^{^{T}}&=&\lambda^{^{T}}_p 
      \COSPHI \; \frac{d\varphi}{2 \pi}      
      \sqrt{\frac{4\Mp x}{Sy}\left(1-y-\frac{\Mp Q^2}{S^2}\right)},
\label{facts}
\ea
and $S^{^{L,T}}_{g1-g5}(y,Q^2)$
are kinematical factors which obey
a compact explicit form similar to~(\ref{sunpol}), see ref.~\cite{ph}.

\noindent
The square root in~(\ref{facts})
is related to the electron scattering angle $\theta_2$
\ba
\sqrt{\frac{4\Mp x}{Sy}\left(1-y-\frac{\Mp Q^2}{S^2}\right)}
      =\frac{1-y}{y}\sin\theta_2.
\ea
\noindent
The angle $\varphi$ is an azimuthal angle between transverse spin 
vectors and reaction plane.
The nucleon polarization vector in~(\ref{spin}) was taken as
\ba
\vec{n}=\lambda^{^L}_p\frac{\vec{k}_1}{|\vec{k}_1|}
\ea
for the longitudinal case, and as
\ba
\vec{n}=\lambda^{^T}_p {\vec{n}}_{\perp}
\ea
for the transverse case
where ${\vec{n}}_{\perp}$ satisfies
\ba
\vec{k_1}.{\vec{n}}_{\perp}=0.
\ea

The expressions~(\ref{unpol}) and~(\ref{polar}) possess
the same factorization property as leptonic tensor~(\ref{lnrt1}) does. As a
consequence of it, the SF's combine with the $\gamma$ and $Z$ propagators and
coupling constants and
factor out from the universal kinematic factors,
$S^{^{U,L,T}}_{F_i,g_j}$, which are simple functions
of the two independent invariants, taken
as $y$ and $Q^2$ for definiteness.

\smallskip

The SF's  $F^{^{J_1 J_2}}_i$ and  $g^{^{J_1 J_2}}_i$
enter actually in only two combinations which are due to only two 
factorizing scalar structures, $L_{_S}$ and $L_{_A}$, in~(\ref{lnrt1}). 
They are sometimes called
{\it generalized} or {\it combined} SF's and read
\ba
{\cal  F }_{1,2}(x,Q^2)
 &=& Q^2_{e} F^{\gamma\gamma}_{1,2}(x,Q^2) \nll
&&+ 2 |\Qe|
 \left( v_e -\pe \lambda_e a_e \right)\chi(Q^2) F^{\gamma Z}_{1,2}(x,Q^2) 
\nll
      &&+
 \left( v^2_e + a^2_e - 2\pe\lambda_e v_e a_e \right) \chi^2(Q^2)
                                                  F^{ZZ}_{1,2}(x,Q^2),
\nll
{\cal F }_{3}(x,Q^2)   &=&
      \Pe\Biggl\{
 2|\Qe|\left( a_e -\pe\lambda_e v_e \right) \chi(Q^2)
                                                  F^{\gamma Z}_3(x,Q^2)  
\nll
      &&+ \left[2 v_e a_e -\pe\lambda_e
          \left( v^2_e + a^2_e\right)\right] \chi^2(Q^2)
                                                  F^{ZZ}_3(x,Q^2)\Biggr\},   
\nll
{\cal G }_{1,2}(x,Q^2) &=& \Pe \Biggl\{
 -Q^2_{e}\pe \lambda_e  g^{\gamma \gamma}_{1,2}(x,Q^2)              \nll
&&   + 2 |\Qe| \left( a_e -\pe\lambda_e v_e \right)\chi(Q^2)
                                         g^{\gamma Z}_{1,2}(x,Q^2)
 \nll
  && +  \left[2 v_e a_e -\pe \lambda_e
         \left( v^2_e + a^2_e \right)\right] \chi^2(Q^2)
                                            g^{ZZ}_{1,2}(x,Q^2)
                                            \Biggr\},  
\nll
{\cal G }_{3,4,5}(x,Q^2) &=&
              2 |\Qe|\left( v_e -\pe \lambda_e a_e \right) \chi(Q^2)
                                        g^{\gamma Z}_{3,4,5}(x,Q^2)
 \nll
  && +  \left( v^2_e + a^2_e - 2\pe\lambda_e v_e a_e \right)
                                         \chi^2(Q^2)
                                        g^{ZZ}_{3,4,5}(x,Q^2).
\label{genstf}
\ea

Eqs.~(\ref{unpol})--(\ref{genstf}) represent the complete set of formulae
for the unpolarized and 
polarized DIS in the Born approximation. Now we turn to the
first order QED RC's within MI approach.

\subsubsection{Radiative process
 $ep \rightarrow e X \gamma$}
For the description of the radiative process $ep \rightarrow e X\gamma$,
one has to distinguish 
{\it leptonic} and {\it hadronic} variables:
\ba
   q_l &=& k_1 - k_2,  \qquad
   q_h\;=\;p'  - p,    \nll
   \Ql &=& -q_l^2,     \qquad   \quad
   \Qh\;=\;-q_h^2,     \nll\
   \yl &=& \frac{2p.q_l}{S}, \qquad \;\;\;
   \yh\;=\;\frac{2p.q_h}{S}. 
\label{radinv}
\ea
Four invariants, $\yl,\;\Ql,\;\yh,\;\Qh,$ together with an azimuthal angle
$\varphi_k$ varying from $0$ to $2\pi$ (see~\cite{mi} for a complete
description of the kinematics of the process $ep \rightarrow eX\gamma$),
form a complete set of five independent kinematic variables.

  The differential cross-section for the scattering of polarized electrons
off polarized protons, originating from the four bremsstrahlung diagrams
(for both $\gamma$ and $Z$-boson exchanges)
has a form similar to~(\ref{sborn2})
\ba
\frac{\ds d\sigma_{_{\mr BREM}}}{\ds d\xl d\yl}
=  { 2  \alpha^3 }
\int d\yh d\Qh \frac{1}{\QhS}
\Biggl[\frac{1}{2\pi}\frac{d\varphi_k}{{\SLQ}}
\frac{S\yl}{4}\Biggl(L^{\mu\nu}_{_{\small rad}} W_{\mu\nu} \Biggr)\Biggr],
\label{brem2}
\ea
with
\bq
\LQ\;=\;S^2 \ylS+4 M^2 \Ql.
\eq
Here $W_{\mu\nu}$ is given by the same 
formulae~(\ref{hadten})--(\ref{spin}) as for the Born case but now 
with all 4-momenta acquiring an index $h$, which stands for {\it hadronic}
variables. This is a property of the MI approach.
The quantity $L^{\mu\nu}_{_{\small rad}}$ denotes 
the {\bf leptonic radiative tensor},
an analog of the Born leptonic tensor~(\ref{lnrt1}), but
for four bremsstrahlung diagrams. Its explicit form
is presented in~\cite{ph}. It does not exhibit such a simple
factorizing structure as~(\ref{lnrt1}), see below.
The unpolarized cross-section reads
\ba
\frac{\ds d\sigma^{^{U}}_{_{\mr {BREM}}}}{\ds d\xl d\yl}
=
  2  \alpha^3 \int d\yh d\Qh \frac{1}{\QhS}
\sum_{i=1}^{3} \;
  S^{^U}_{i}(\yl,\Ql,\yh,\Qh) {\cal F}_i(\xh,\Qh).
\label{ubrem}
\ea
The explicit form of the kinematic factors $S^{^U}_{i}$
is given by eqs.~(3.14)-- (3.16) of ref.~\cite{mi}. They are analogs of
the factors~(\ref{sunpol}) for the case of bremsstrahlung. 
Due to this they
are functions of four invariant variables~(\ref{radinv})
(they are assumed to be
integrated over the angle $\varphi_k$).
As is seen from~(\ref{ubrem}), the factorization for the three
generalized SF's is fulfilled for the unpolarized  cross-section.

The polarized DIS bremsstrahlung cross-sections have
a more complicated structure:
\ba
\frac{\ds d\sigma^{^{L,T}}_{_{\mr {BREM}}}}{\ds d\xl d\yl}
&=&
  2 \alpha^3 f^{^{L,T}} \int d \yh \frac{d\Qh}{\QhS} \Biggl\{
\sum_{i=1}^{5}\;S^{^{L,T}}_{gi}(\yl,\Ql,\yh,\Qh) {\cal G}_i(\xh,\Qh)
\nll
&&+\lambda_e 2 \Me \frac{(B_1,1)}{C^{3/2}_1} \Biggl[
  \sum_{i=1}^{2}\Biggl(
   {\cal S}^{^{L,T}}_{vi}(\yl,\Ql,\yh,\Qh){\cal G}^v_i(\xh,\Qh)
\nll
&&\hspace{3.30cm}
  +{\cal S}^{^{L,T}}_{ai}(\yl,\Ql,\yh,\Qh){\cal G}^a_i(\xh,\Qh)\Biggr)
\nll
&&\hspace{2.20cm}
 +\sum_{i=3}^{5}
   {\cal S}^{^{L,T}}_{zi}(\yl,\Ql,\yh,\Qh){\cal G}^z_i(\xh,\Qh)
\Biggr]
\Biggr\}.
\label{pbrem}
\ea

We note that the first sum in~(\ref{pbrem}) exhibits
the same factorization property as~(\ref{ubrem}), but now for five
polarized SF's. There also appear seven new
generalized SF's, ${\cal G}^{v,a,z}_i(\xh,\Qh)$,
and seven associated kinematic factors, ${\cal S}^{^{L,T}}_{vi,ai,zi}$.
All these non-factorizable terms are proportional to $\lambda_e$
and $m^2$.
These contributions turn out to be rather important
since after one integration more they yield terms of ${\cal O}(1)$
\footnote
{In the notation of ref.~\cite{mi}, these are terms of
${\cal O}(m^2/z^2_1)$, which are known to give a non-negligible
contribution in complete 
${\cal O}(\alpha)$ calculations.}.  

The additional generalized SF's are:
\ba
{\cal G}^v_{1,2}(x,Q^2) &=& Q^2_{e}g^{\gamma \gamma}_{1,2}(x,Q^2)
\\
 &&           + 2|\Qe|v_e\chi(Q^2) g^{\gamma Z}_{1,2}(x,Q^2)
              + v^2_e\chi^2(Q^2)g^{ZZ}_{1,2}(x,Q^2),             \nll
{\cal G}^a_{1,2}(x,Q^2) &=& a^2_e\chi^2(Q^2)g^{ZZ}_{1,2}(x,Q^2),  \nll
{\cal G}^z_{3,4,5}(x,Q^2) &=& |\Qe|a_e\chi(Q^2) g^{\gamma Z}_{3,4,5}(x,Q^2)
              + v_e a_e \chi^2(Q^2) g^{ZZ}_{3,4,5}(x,Q^2).
\nonumber
\label{addgenstf}
\ea

All kinematic factors $S^{^{L,T}}_{gi}$
and ${\cal S}^{^{L,T}}_{vi,ai,zi}$ 
are of comparable complexity to those for the unpolarized DIS.
They all were explicitly derived in~\cite{ph}.
The expressions for $B_{1,2}$ and $C_{1,2}$ 
are given in~\cite{mi}, eqs. (A.30)--(A.31).

\subsubsection{The net radiative correction}

In all figures we show the
dimensionless radiative correction factor:
\ba
\delta^{k}_{l} \equiv \delta^{k}(\xl,\yl)=
 \frac{d^{2}{\sigma}^{k}_{_{\mr {RAD }}}/d\xl d\yl}
      {d^{2}{\sigma}^{k}_{_{\mr {BORN}}}/d\xl d\yl}-1,
\label{delta}
\ea
where
$ d^{2}{\sigma}^{k}_{_{\mr {BORN}}}$
is the Born cross-section for DIS and
$ d^{2}{\sigma}^{k}_{_{\mr {RAD }}}$ is the
{\it radiatively corrected} cross-section.
The index $k$ runs over {\it unpolarized}, {\it longitudinal} and
{\it transverse} configurations.
The cross-section $ d^{2}{\sigma}^{k}_{_{\mr {RAD }}}$
is usually presented as the sum of two terms:
\ba
\frac{d^{2}{\sigma}^{k}_{_{\mr {RAD}}}}{d\xl d\yl}
&=&
\left[1+\frac{\alpha}{\pi}\;\delta_{_{\mr {VR}}}(\xl,\yl)\right]
\frac{d^{2} {\sigma}^{k}_{_{\mr {BORN}}}}{d\xl d\yl}
+\frac{d^{2}{\sigma}^{k}_{_{\mr {R}}}}{d\xl d\yl}.
\label{sirad}
\ea
The first term contains the {\it universal, factorized correction}, 
originating from the {\it vertex}
diagram and an IR-divergent part of the
bremsstrahlung contribution.
In {\em leptonic} variables it is given by eq.~(4.45) of~\cite{mi}.
The second {\it non-universal, non-factorized} term
originates from the rest of bremsstrahlung contributions,
which are free of IR-divergences by construction:
\ba
\frac {d^2 \sigma^{k}_{_{\mr R}}}{d \xl d \yl}
= 2\alpha^3
     \int d\yh d\qh \Biggl[\frac{1}{Q^4_h} \sum_{i}  
{\cal S}^{k}_{i}(\yl,\ql,\yh,\qh)
     {\cal F}^{k}_{i}(\xh,\qh)
\nll
\hspace{1cm}-\frac{1}{Q_{l}^4} \sum_{i'}
{\cal S}^{k}_{i'_{\mr {BORN}}}(\yl,\ql)
     {\cal L}^{\mr {IR}}(\yl,\ql,\yh,\Qh){\cal F}^{k}_{i'}(\xl,\ql)
\Biggr],
\label{subtr}
\ea
where ${\cal L}^{\mr {IR}}(\yl,\Ql,\yh,\Qh)$ is given by eq.(5.4) of~\cite{mi}.

In~(\ref{subtr}),
the indices $i,i'$ run over the set of kinematical
factors and generalized SF's $\cal{F}$ or $\cal{G}$ relevant to the index $k$.
We note that all ``additional'' terms in~(\ref{pbrem})
of ${\cal O}(m^2)$ are infrared finite.
Therefore they need not be subtracted in~(\ref{subtr}). Due to this 
one has two different indices $i,i'$ running in different limits. 

The formulae of this subsection, together with all kinematical factors
being not given here, present a complete set of formulae for the MI
approach to the RC's for polarized DIS.
Here the presentation follows the spirit of 
the review~\cite{mi}.

\subsection{Leading Logarithmic Approximation}

In the leading logarithmic approximation (LLA),
the \oa\ corrections consist of three incoherent contributions due to
initial and final state radiations (ISR and FSR)~\cite{ll0}-\cite{ll2} 
and the Compton peak~\cite{mots},~\cite{compton},~\cite{ll3}
\ba
\frac{d^2 \sigma_{_{\mr {RAD}}}}{dxdy}
&=&
\frac{d^2 \sigma_{i}}{dxdy}
+
\frac{d^2 \sigma_{f}}{dxdy}
+
\frac{d^2 \sigma_{_{\rm{COMP}}}}{dxdy}.
\label{lla3}
\ea
The ISR and FSR cross-sections have a similar generic structure:
\ba
\frac{d^2 \sigma^{k,a}_{i,f}}{dxdy}
&=&
\frac{\alpha}{2\pi}
\left( \ln\frac{Q_a^2}{m^2}-1 \right)
\int\limits_{0}^{1}dz 
\frac{1+z^2}{1-z}
\nll
&&\times
\Biggl\{\theta(z-z_0){\cal J}\left.
\frac{d^2
  \sigma^{k}_{_{\rm{BORN}}}}{dxdy}\right|^{a;i,f}_{x={\hat x},
 y={\hat y}, S={\hat S}}
-\frac{d^2 \sigma^{k}_{_{\rm{BORN}}}}{dxdy}\Biggr\},
\nll
{\hat x} &=& \frac{{\hat Q}^2}{{\hat y}{\hat S}},
\quad
{\cal J} \; \equiv \; {\cal J}(x,y,Q^2) =
\left|\frac{\partial({\hat x},{\hat y})}{\partial(x,y)}\right|.
\label{llaif}
\ea
The lower integration boundary  $z_0$ derives from the conditions
\ba
{\hat x}(z_0) &\leq& 1, \qquad
{\hat y}(z_0) \; \leq\;  1.
\label{llascal}
\ea
Here the new index $a$ stands for 
the different types of measurements, 
for which 
the definitions of the ${\hat x},\;{\hat y},\;
{\hat S}$, as well as of the $z_0$, are known to be different
(see for example~\cite{he}). 
Formulae of similar structure are known in the
second order LLA, \oaal~\cite{kripf}.

For leptonic variables all the Compton 
contributions are known~\cite{ph}:
\ba
\frac{d^2 \sigma_{_{\rm{COMP}}}^{^U}}{d\xl d\yl}
&=&
\frac{\alpha^3}{Sx^2_l } \frac{Y_{+}}{y_{l_1}}
  \int \limits_{x_l}^{1} \frac{d\xh}{x_h}
  \int\limits^{(Q^2_h)^{\rm{\small{max}}}}_{(Q^2_h)^{\rm{\small{min}}}}
  \hspace{-2mm}
  \frac{d\Qh}{\Qh}
  \Biggl[Z_{+}F^{\gamma\gamma}_2(\xh,\Qh)
\nll
&&\hspace{5cm}
-\left(\frac{\xl}{\xh}\right)^2 F^{\gamma\gamma}_{_L}(\xh,\Qh)\Biggr],
\label{compt_u}
\nll
\frac{d^2 \sigma_{_{\rm{COMP}}}^{^L}}{d\xl d\yl}
&=&
  \left(- 2 \lambda_e \lambda^{^L}_p \right)
  \frac{\alpha^3}{S x^2_l }\frac{Y_-}{\yll}
  \int \limits_{x_l}^{1} {d\xh}
  \int\limits^{(Q^2_h)^{\rm{\small{max}}}}_{(Q^2_h)^{\rm{\small{min}}}}
\hspace{-2mm}
  \frac{d\Qh}{\Qh}
  Z_{-}g^{\gamma\gamma}_1(\xh,\Qh),
\nll
\frac{d^2 \sigma_{_{\rm{COMP}}}^{^T}}{d\xl d\yl}
&=&
  \left(- 2 \lambda_e \lambda^{^T}_p \right)
  \frac{\alpha^3}{S x^2_l }
  \cos \varphi \frac{d\varphi}{2 \pi}
  \frac{y_l}{y^2_{l_1}}
  \sqrt{\frac{4 M^2 x_l}{S y_l} \left(\yll-\frac{M^2 x_l y_l}{S} \right)}
\nll
&&\times
\int \limits_{x_l}^{1} {d\xh}
\int\limits^{(Q^2_h)^{\rm{\small{max}}}}_{(Q^2_h)^{\rm{\small{min}}}}
\hspace{-2mm}
  \frac{d\Qh}{\Qh}
  \Biggl\{
  \left( Y_{-} - y_l z \right) z
  g^{\gamma\gamma}_1(\xh,\Qh) 
\nll
& &\hspace{1.5cm} 
+ 2  \left[ Y_{+}\left(1-z \right)
 + y_{l_1} \right]
  g^{\gamma\gamma}_2(\xh,\Qh),
  \Biggr\}.
\label{compt}
\ea
where
\ba
Y_{\pm}&=& 1 \pm \yllS,
\quad
Z_{\pm}\;=\; \left[ 1 \pm \left(1-z\right)^2\right],
\nll
\yll &=& 1 - \yl,
\qquad
z\;=\;\frac{\xl}{\xh}.
\label{yz}
\ea

The LLA formulae are remarkably compact. To derive 
the ISR and FSR contributions one has to know only the Born
cross-section. No more complex are also the relations for 
the Compton peak contributions.
A natural question arises: How precise are they as compared to complete 
${\cal O}(\alpha)$ calculations? We will present some
figures with comparisons of LLA and complete calculations in section 3.

\subsection{QPM Approach, EWRC}
The only way to go beyond {\it leptonic} corrections
is to give up the MI approach in favour of
complete \oa\ calculations within
the framework of the quark-parton model (QPM) approach
where one can access the following RC's:
\begin{itemize}
\item[1.]
The QED RC's to the leptonic current;
\item[2.]
The QED RC's to the quark current -- a model of hadronic RC's;
\item[3.]
The interference of lepton and quark bremsstrahlung
together with the corresponding $\gamma \gamma, \gamma Z, \gamma W$
boxes;
\item[4.]
The electroweak radiative corrections (EWRC).
\end{itemize}

If identical SF's are chosen,
the QPM
leptonic current QED corrections (1.) should agree exactly
with those calculated in the MI
approach. However,
there is no access in the MI approach to the 
corrections~(2.),~(3.), and~(4.).
The EWRC~(4.) are usually taken into account using the language 
of {\it effective weak couplings}.

{\tt HECTOR 1.00}~\cite{he} contains two QPM-based branches with
complete ${\cal O}(\alpha)$ QED and EWRC's to:
\begin{itemize}
\item
 NC and CC DIS in {\em leptonic} variables~\cite{qp};
\item
 NC DIS in {\em mixed} variables~\cite{mx}.
\end{itemize}

\section{Numerical Results}
\begin{figure}[tb]
\center
\vspace{-5mm}
\mbox{\epsfig{file=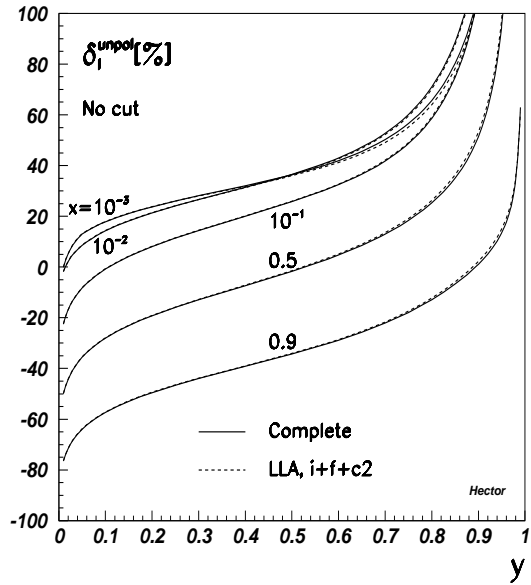,height=9cm,width=8cm}}
\vspace{-5mm}
\caption{A comparison of complete and LLA RC's at HERA collider 
kinematic regime
for NC {\it unpolarized} DIS in {\it leptonic} variables.}
\label{fig1}
\end{figure}
\begin{figure}[tb]
\center
\vspace{-8.mm}
\mbox{\epsfig{file=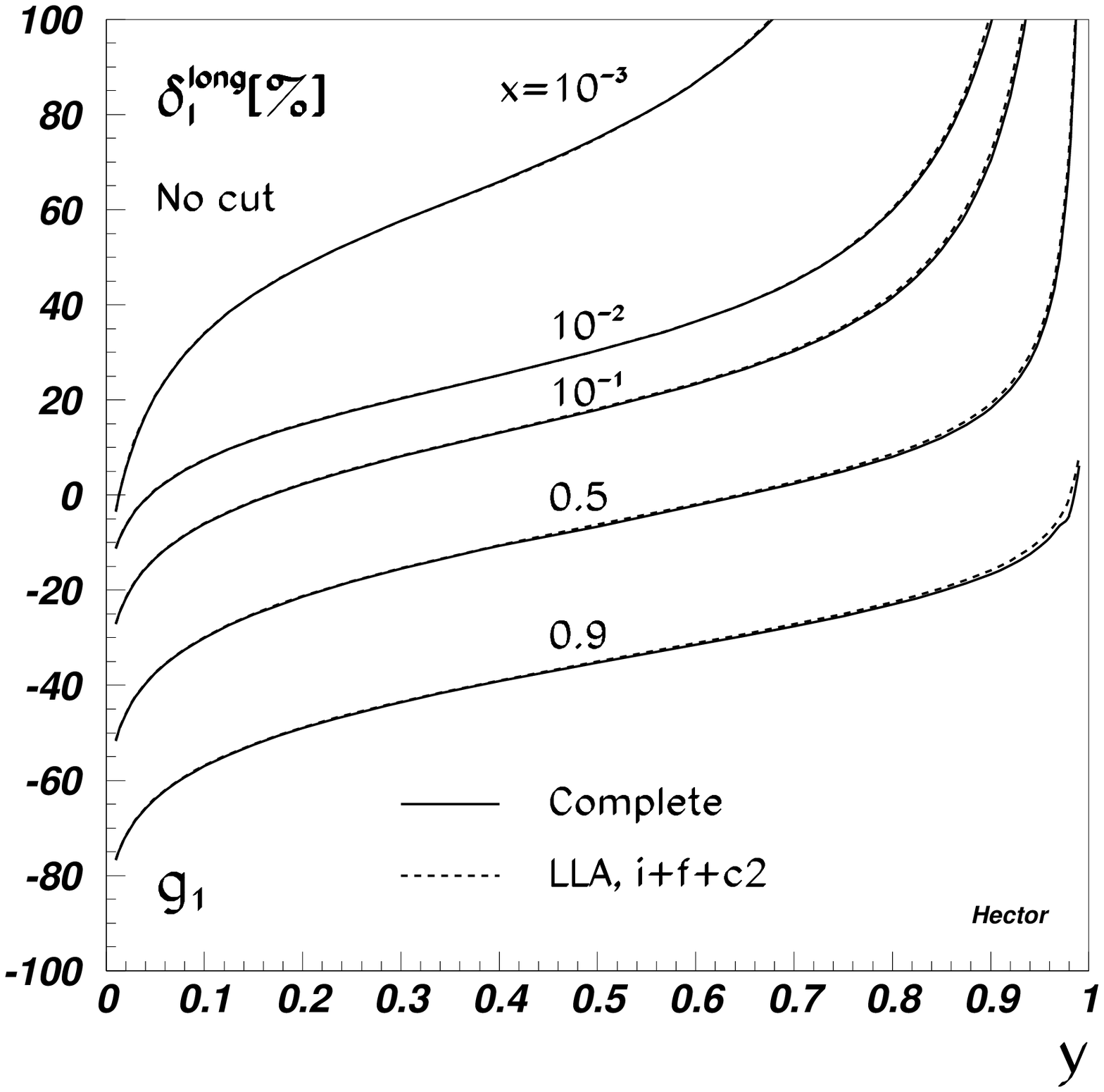,height=9cm,width=8cm}}
\vspace{-8.mm}
\caption{The same as Figure 1 but for {\it longitudinal} DIS.}
\label{fig2}
\end{figure}
In this section we present some numerical results obtained with 
an upgraded version of the {\tt HECTOR} package~\cite{h2} and present 
an updated comparison with the results obtained by the code 
{\tt POLRAD15}~\cite{pr}.

For a brief description of main features of these codes 
as well as for some numerical results illustrating
the comparison between LLA and complete ${\cal O}(\alpha)$
calculations, and for a first comparison of these two codes 
we refer the reader 
to~\cite{db} and~\cite{hy}.

In all numerical calculations
we used the CTEQ3M parametrization~\cite{3m}
for the unpolarized SF's and the GRSV'96 parametrization~\cite{96}
for the polarized SF's. 

\subsection{New Results of LLA/Complete Comparison}

In figures~\ref{fig1} and~\ref{fig2} a comparison of the RC
factors~(\ref{delta}) is shown. For the calculations we used the 
${\cal O}(\alpha)$ QED formulae presented in subsections 2.2 and  
2.3 of this report. In the LLA calculations, all the three 
contributions~(\ref{lla3}) were used. We note that taking into
account the Compton peak contribution in the form of the two-fold
integral~(\ref{compt}) with a tuned upper limit $(Q^2_h)^{\rm{max}}$
(see~\cite{ph} for details)
improves substantially the agreement as compared to the case when
only initial and final state RC~(\ref{llaif}) were considered.
  
An agreement at the same level of precision persists even if a cut
on the invariant mass of the final hadronic state, $M^2_h$, or on the 
transfer momentum, $Q^2_h$, of the order of $100$~GeV$^2$ is imposed.
    
We would like to warn the reader, however, that taking into account LLA alone
is {\em not} fully sufficient in all cases. In particular, at HERMES energies
the agreement becomes poorer, see figures~\ref{fig3} and~\ref{fig4} below,
and even worse when rather loose cuts are imposed.

\subsection{An Updated Comparison of
{\tt HECTOR~1.11} and {\tt POLRAD15} Results}

This comparison, like the previous one, was done for the kinematic range
of HERMES, for the leptonic measurement of polarized DIS on a
proton target
both for longitudinal and transverse orientations of the proton spin.
Only the $\gamma$ exchange diagrams and 
the first order QED RC's were retained.

\begin{figure}[tb]
\center
\vspace{-8mm}
\mbox{\epsfig{file=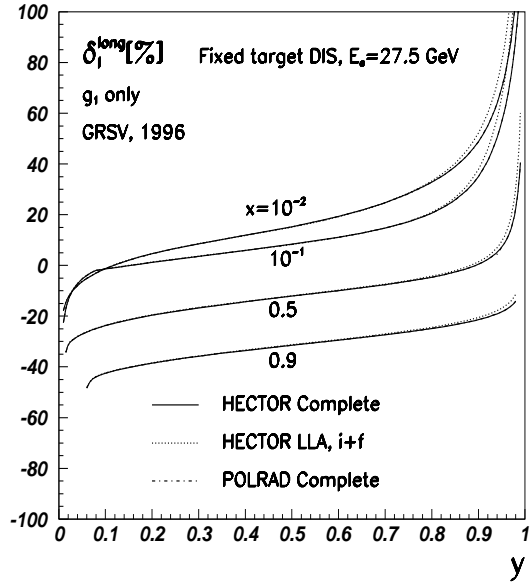,height=9cm,width=8cm}}
\vspace{-8mm}
\caption{A comparison of RC's calculated by {\tt HECTOR} and {\tt POLRAD}
for NC {\it longitudinal} DIS in {\it leptonic} variables.}
\label{fig3}
\end{figure}

\begin{figure}[tb]
\center
\vspace{-8mm}
\mbox{\epsfig{file=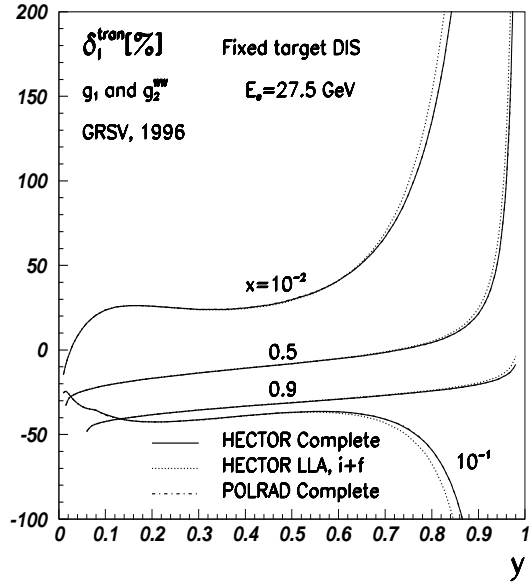,height=9cm,width=8cm}}
\vspace{-8mm}
\caption{The same as Figure 3 but for {\it transverse} DIS. For $g_2$
the Wandzura--Wilczek relation~\cite{ww} was used.}
\label{fig4}
\end{figure}

Figures~\ref{fig3} and~\ref{fig4} update corresponding figures
of~\cite{db} and ~\cite{hy}.
These figures, together with a figure from ref.~\cite{hy} 
for the unpolarized DIS,
demonstrate a very good agreement of the results
of the ``tuned'' (i.e. with exactly the same, simplified input) comparison between
{\tt HECTOR 1.11} and {\tt POLRAD15}. 
This does not replace future comparisons in the real experimental applications.
The previously 
registered small disagreement for the 
polarized cases
for low $x$ and high $y$ was due to the omission
of terms of~${\cal O}(m^2)$ in eq. (\ref{pbrem})
in~\cite{db} and~\cite{hy}.

From figures~\ref{fig3} and~\ref{fig4} one can also see how well the
LLA and complete calculations do agree at HERMES energies.

\section{RC for tagged photon DIS}

The interest to DIS with {\it tagged photons} arose recently.
The H1 and ZEUS collaborations collected samples of DIS events in which a
photon is observed in the so-called backward luminosity tagger with a typical
angular acceptance of 0.5~mrad around the beam axis.
Although the present statistics is limited to several thousand events,
it will largely improve with more HERA data coming. This is the reason
why the RC to this sample have to be calculated at the percent level of precision.
The relevant DIS Born-level cross-section, instead of~(\ref{unpol}),
is described by a three fold-differential expression
\ba
 \frac{d^3 \sigma_{\rm{brem}}}
{d\xl d\yl d E_{\gamma}} 
=
\frac{2\alpha^3}{\pi} \yl
\int  d\cos\theta_\gamma
 \int d\varphi_\gamma
\frac{E_{\gamma}}{Q^4_h}
\sum^3_{i=1} S^{^{U}}_{i} {\cal F}_i,
\label{lowest}
\ea 
where the integration is performed over the angular range covered by the
photon tagger.
In~(\ref{lowest}) the kinematic factors $S^{^{U}}_{i}$,
contrary to~(\ref{ubrem}), are understood to be completely
differential in five kinematic variables. The latter are chosen as
$\xl,\;\yl$, and $\theta_\gamma,\;\varphi_\gamma,\;E_{\gamma}$ in the
laboratory frame, where the photon variable cuts are defined.
We note that the usual definitions of $\xl$ and
$\yl$~(\ref{radinv}) are used in this section, i.e. they are not
recalculated using the {\it reduced} electron beam energy.

In a recent paper~\cite{tp}, we performed a detailed calculation of
the Born cross-section~(\ref{lowest}) and an evaluation of the RC to it.
The main idea is to combine the MI approach for
the description of the Born cross-section~(\ref{ubrem}) (the DIS bremsstrahlung
is the Born-level process in the problem under consideration)
with the LLA for the description of ISR QED corrections~(\ref{llaif}).

\begin{figure}[tb]
\center
\vspace{-8mm}
\mbox{\epsfig{file=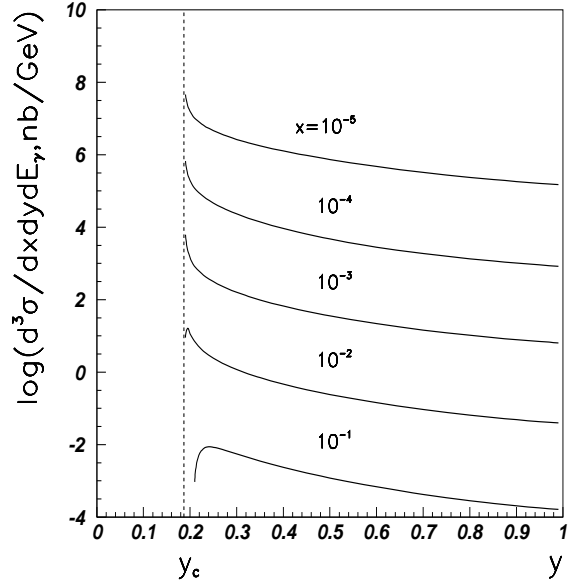,height=9cm,width=8cm}}
\vspace{-8mm}
\caption{The three-fold differential Born DIS cross-section with 
{\it tagged photons} for $E_\gamma=5$ GeV; $y_c=E_\gamma/E_e$, $E_e=27.5$ GeV.}
\label{fig5}
\end{figure}

\begin{figure}[tb]
\center
\vspace{-8mm}
\mbox{\epsfig{file=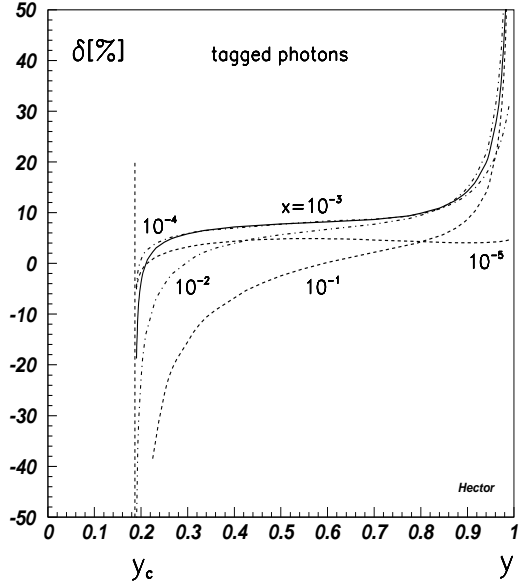,height=9cm,width=8cm}}
\vspace{-8mm}
\caption{The RC to DIS cross-section with {\it tagged photons}.}
\label{fig6}
\end{figure}

In figures~\ref{fig5} and~\ref{fig6} we show the Born
cross-section and the RC for $E_\gamma=5$ GeV, the peak value of
the distribution of tagged DIS events.

The RC exhibits nice properties: it shows a typical behaviour in soft 
and hard bremsstrahlung corners in $y$ and is quite flat in between. 
In the plateau region, its value is limited within a $\pm 10\%$
band. The RC grows slightly with increasing $E_\gamma$, e.g. for
$E_\gamma=10$~GeV the plateau behaviour becomes less pronounced
and for reasonable $\xl$ and $\yl$ (e.g. $\yl < 0.9$) its
value is limited within a $+5\%,+25\%$ interval.

The fact that the RC's for DIS with tagged photons are
not so big gives reasons to trust a simplified approach as used here.
However, a complete calculation of ${\cal O}(\alpha)$ RC's to the DIS
bremsstrahlung cross-section
seems to be still an important physical task in view of high statistics
data to be taken at HERA in the coming years.


\subsection*{Acknowledgments}
The authors are very much obliged
to I.~Akushevich for fruitful collaboration
on the comparison with {\tt POLRAD} and for important remarks.
We are thankful to A.~Arbuzov 
for an enjoyable common work within the {\tt HECTOR} project.
We are grateful to
M.~Klein for  reading of the manuscript.

\end{document}

%% file: lk_newcommand.tex
\newcommand{\bq}{\begin{equation}}
\newcommand{\eq}{\end{equation}}
\newcommand{\ba}{\begin{eqnarray}}
\newcommand{\ea}{\end{eqnarray}}

\newcommand{\mr}{\rm}
\newcommand{\ds}{\displaystyle}

\newcommand{\nll}{\nonumber \\}

\newcommand {\Ql  }{\mbox{$Q^2_{l}    $}}

\newcommand {\Qh  }{\mbox{$Q^2_{h}    $}}
\newcommand {\QhS }{\mbox{$Q^4_{h}    $}}

\newcommand {\yl  }{\mbox{$y  _{l}    $}}
\newcommand {\ylS }{\mbox{$y^2_{l}    $}}
\newcommand {\xl  }{\mbox{$x  _{l}    $}}
\newcommand {\yll }{\mbox{$y  _{l_1}  $}}
\newcommand {\yllS}{\mbox{$y^2_{l_1}  $}}
\newcommand {\yh  }{\mbox{$y_{h}      $}}

\newcommand {\xh  }{\mbox{$x  _{h}    $}}

\newcommand {\Me  }{\mbox{$m^2        $}}
\newcommand {\Mp  }{\mbox{$M^2        $}}

\newcommand {\SLQ }{\mbox{$\sqrt{\lambda_q}  $}}

\newcommand {\Qe   }{\mbox{$Q_{e}   $}}

\newcommand {\COSPHI }{\mbox{$\cos{\varphi}                $}}

\newcommand{\pspin}{{\it s}           }
\newcommand{\Pe   }{\mbox{$p_e      $}}
\newcommand{\pe   }{\mbox{$p_e      $}}

\newcommand{\LQ    }{\mbox{$     {\lambda_q}   $}}

%% file: Cracow_proc.bbl
\newpage

\begin{thebibliography}{99}
%
\bibitem{ph}
D.~Bardin, J.~Bl\"umlein, P.~Christova and L.~Kalinovskaya,
DESY 96--189.
%
\bibitem{tp}
D.~Bardin, L.~Kalinovskaya and T.~Riemann,
DESY 96--213.
%
\bibitem{db}
D.~Bardin, to appear in {\it Proceedings of 28th International Conference on
High Energy Physics, 25-31 July 1996, Warsaw, Poland.}
%
\bibitem{Bhabha}
S. Jadach, O. Nicrosini (conveners)  et al.,
in: G. Altarelli, T. Sj\"ostrand, F. Zwirner, eds.,
{\em Proceedings of the Workshop on Physics at LEP~2},
CERN Yellow Report CERN 96-01 (1996), {\bf Vol.~2}, p. 229.
%
\bibitem{LEP2}
D. Bardin, R. Kleiss (conveners) et al.,
{\it ibid.} {\bf Vol.~2}, p. 3.
%
\bibitem{hr}
A.~Kwiatkowski, H.-J.~M\"ohring and H.~Spiesberger, 
in: {\it Proceedings of the
Workshop on Physics at HERA, Oct. 29--30, 1991, 
Hamburg (DESY,
Hamburg, 1992)}, W.~Buchm\"uller and G.~Ingel-

\pagebreak

man (eds.),
{\bf Vol.~3}, p. 1294;\Journal{\CPC}{69}{155}{1992}.
%
\bibitem{lf}
S.~Jadach and W.~Placzek, 
{\it ibid.} {\bf Vol.~3}, p. 1330.
%
\bibitem{kr}
H.~Anlauf, H.D.~Dahmen, P.~Manakos, T.~Mannel and T.~Ohl,
{\it ibid.} {\bf Vol.~3}, p. 1311.
%
\bibitem{hl}
J.~Bl\"umlein, {\it ibid.} {\bf Vol.~3}, p. 1270.
%
\bibitem{tr}
A.~Akhundov, D.~Bardin, L.~Kalinovskaya and T.~Riemann,
{\it ibid.} {\bf Vol.~3}, p. 1285.
%
\bibitem{fr}
{\sc Fortran} code {\tt FERRAD35}
used by SMC and based on ref.~\cite{mots}.
%
\bibitem{mots}
L.W.~Mo and Y.S.~Tsai, {\it Rev. Mod. Phys.} {\bf 41} (1969) 205.
%
\bibitem{ar}
G.~Montagna, O.~Nicrosini and L.~Viola,
in: {\it the same Proceedings as in ref.}~\cite{hr}, {\bf Vol.~3}, p. 1267.
%
\bibitem{pr}
I.~Akushevich and N.~Shumeiko, \Journal{\JPG}{20}{513}{1994};
                               {\it Yad. Fiz.} {\bf 58} (1995) 507.
%
\bibitem{he}
A.~Arbuzov, D.~Bardin, J.~Bl\"umlein, L.~Kalinovskaya and T.~Riemann,
\Journal{\CPC}{94}{128}{1996}.
%
\bibitem{da}
G.~Passarino,
\Journal{\CPC}{97}{261}{1996}.
%
\bibitem{to}
T.~Ohl, {\it these Proceedings}; hep-ph/9610349.
%
\bibitem{mu}
D.~Bardin and L.~Kalinovskaya, {\sc Fortran} code
${\mu}${\bf e}{\it la} 1.00.
%
\bibitem{h2}
A.~Arbuzov, D.~Bardin, J.~Bl\"umlein, P.~Christova, L.~Kalinovskaya and T.~Riemann,
{\sc Fortran} code {\tt HECTOR 1.11}.
%
\bibitem{mi}
A.~Akhundov, D.~Bardin, L.~Kalinovskaya and T.~Riemann, 
\Journal{\FdP}{44}{373}{1996}.
%
\bibitem{blukoch}
J.~Bl\"umlein and N.~Kochelev, \Journal{\PLB}{381}{296}{1996}.
%
\bibitem{ll0}
A. De Rujula, R.~Petronzio and A. Savoy--Navarro,
{\it Nucl. Phys.} 
{\bf B154} (1979) 394; \\

\pagebreak

M.~Consoli and M.~Greco,
{\it Nucl. Phys.} {\bf B186} (1981) 519;\\
E. Kuraev, N.~Merenkov and V. Fadin,
{\it Sov. J. Nucl. Phys.} {\bf 47} (1988) 1009.
%
\bibitem{ll1}
J. Bl\"umlein, \Journal{\ZPC}{47}{89}{1990};
\Journal{\PLB}{271}{267}{1991}.
%
\bibitem{ll2}
I.~Akushevich, T.~Kukhto, {\it Yad. Fiz.} {\bf 52} (1990) 1442;
{\it Acta Phys. Polonica} {\bf B22} (1991) 771.
%
\bibitem{compton}
J. Bl\"umlein, G. Levman and H. Spiesberger,
in: {\em Proceedings of the Workshop on Research
Directions of the Decade, Snowmass 1990}, 
ed. E. Berger (World Sci., Singapore, 1992);
\Journal{\JPG}{19}{1695}{1993}.
%
\bibitem{ll3}
I.~Akushevich, T.~Kukhto and F.Pacheco,
\Journal{\JPG}{18}{1737}{1992}. 
%
\bibitem{kripf}
J.~Kripfganz, H.~M\"ohring and H.~Spiesberger,
\Journal{\ZPC}{49}{501}{1991};\\
J. Bl\"umlein,
\Journal{\ZPC}{65}{293}{1995}. 
%
\bibitem{qp}
D. Bardin, C. Burdik, P. Christova and 
T.~Riemann, \Journal{\ZPC}{42}{679}{1989};
\Journal{\ZPC}{44}{149}{1989}.
%
\bibitem{mx}
D.~Bardin, P.~Christova, L.~Kalinovskaya and T.~Riemann,
\Journal{\PLB}{357}{456}{1995}.
%
\bibitem{hy}
D.~Bardin, J.~Bl\"umlein, P.~Christova and L.~Kalinovskaya,
Preprint  DESY 96--198, September 1996; in: {\it Proceedings of the
  Workshop ``Future Physics at HERA'', G.~Ingelman, A.~De~Roeck,
  R.~Klanner (eds.)}, {\bf Vol.~1}, p. 13; hep-ph/9609399.
%
\bibitem{3m}
H.L. Lai, J. Botts, J. Huston, J.G. Morfin, J.F. Owen, J.W. Qiu,
W.K. Tung and H.~Weerts,
\Journal{\PRD}{51}{4763}{1995}.
%
\bibitem{96}
M.~Gl\"uck, E.~Reya, M.~Stratmann and W.~Vogelsang,
{\it Phys. Rev.} {\bf D53} (1996) 4775.
%
\bibitem{ww}
S.~Wandzura and F.~Wilczek,
{\it Phys. Lett.} {\bf B72} (1977) 195.
\end{thebibliography}
